\documentclass{iau}
\usepackage{graphicx}

\title[IAUS291.~~New results from LOFAR] 
{New results from LOFAR} 

\author[V. Kondratiev et al. ]  
{Vladislav Kondratiev$^1$,\\ on behalf of Ben Stappers$^2$ and the LOFAR Pulsar Working Group}
 
\affiliation{$^1$ASTRON, the Netherlands Institute for Radio  Astronomy, \\ 
Postbus 2, 7990 AA Dwingeloo, The Netherlands \\ email: {\tt vlad.kondratiev@gmail.com} \\[\affilskip] 
$^2$Jodrell Bank Center for Astrophysics, School of Physics and Astronomy, \\ The University of Manchester, Manchester M13 9PL, UK \\
email: {\tt Ben.Stappers@manchester.ac.uk}}

\pubyear{2012}
\volume{291}  
\jname{\mbox{Neutron Stars and Pulsars: Challenges and Opportunities after 80 years}}
\editors{J. van Leeuwen, ed.} 
\begin{document}

\maketitle

\begin{abstract}
The LOw Frequency Array, LOFAR, is a next generation radio telescope with its core in the Netherlands and elements 
distributed throughout Europe. It has exceptional collecting area and wide bandwidths at frequencies from 10 MHz up to 250~MHz. 
It is in exactly this frequency range where pulsars are brightest and also where they exhibit rapid changes in their emission
profiles. Although LOFAR is still in the commissioning phase it is already collecting data of high quality. I will present 
highlights from our commissioning observations which will include: unique constraints on the site of pulsar emission, 
individual pulse studies, observations of millisecond pulsars, using pulsars to constrain the properties of the magneto-ionic 
medium and pilot pulsars surveys. I will also discuss future science projects and advances in the observing capabilities.
\keywords{pulsars: general, telescopes: LOFAR}
\end{abstract}


\firstsection 
\section{Introduction}

LOFAR is an interferometric array of dipole antenna stations distributed over the Netherlands and a few countries in Europe, 
that operates at the very low radio frequencies from 10 to 250 MHz. It consists of 24 core stations with the central
part of about 300~m across occupied by 6 stations, which is called ``Superterp''. The Superterp provides a collecting
area comparable to that of the 100-m Green Bank Telescope with a beam size of $\sim 0.5^{\circ}$ at 140~MHz. 
For the full core with a size of about 2 km across, 
it is already Arecibo-like collecting area and beam size of $\sim 5^{\prime}$. At the moment there are also 9 remote
stations and 8 international stations included in the array. The latter are twice as big as Dutch stations and they can operate 
independently from the rest of the array. They are very powerful telescopes in their own right, each with collecting 
area comparable to that of the 64-m Parkes radio telescope.

LOFAR's frequency range spans from 10 to 250~MHz which is achieved by using two types of dipoles, low-band
antennas (LBA) at 10--90~MHz and high-band antennas (HBA) at 110--250~MHz. The HBA dipoles have bow-tie
shape and grouped into tiles of 16 dipoles each. There are 48 tiles in Dutch stations and 96 in the international
stations. LOFAR operates at the lowest radio frequencies visible from the Earth as at 10~MHz there is a cut-off due 
to ionospheric reflection. Operating at  these low frequencies, LOFAR {\it covers the lowest 4 octaves of the radio window}, 
that makes it a very  unique telescope, the only one working at such low radio frequencies {\it with} huge instantaneous 
fractional bandwidth.

\begin{figure}[tbhp]
\begin{center}
\includegraphics[width=0.75\textwidth]{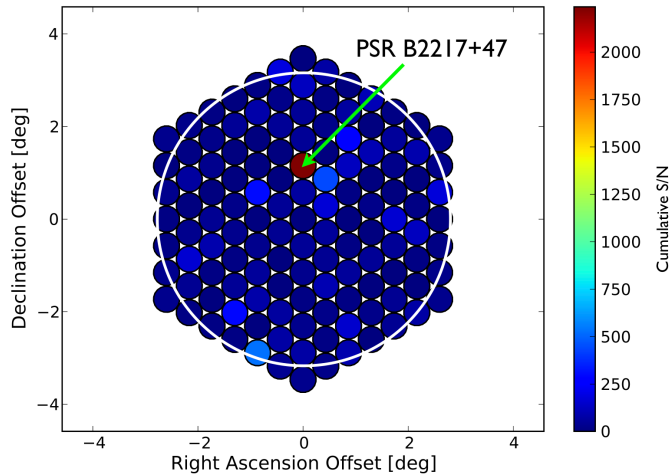}
\caption{Hexagonal pattern of 127 tied-array beams formed around the pulsar B2217+47 using 12 HBA sub-stations
of the Superterp. The beam with the pulsar is marked by the arrow. The cumulative signal-to-noise ratio 
is designated by the color scale. The white circle of about 5 degrees across represents the whole station beam
of the single HBA sub-station.} 
\label{ta127}
\end{center}
\end{figure}

\section{LOFAR Capabilities}

All of the beam-formed modes, capabilities and many commensal results are described in detail in our 
first LOFAR paper \cite[(Stappers et al. 2011)]{sha+11}. Here we
highlight a few of those,
focusing on new commissioning advancements.

{\bf Multiple station beams.} LOFAR is a very flexible, electronically steered aperture array. It is possible to form multiple station beams 
on the sky and observe several pulsars simultaneously (\cite[Hessels et al. 2010]{hsa+10}; \cite[Stappers et al. 2011]{sha+11}). 
This technology will be crucial for the SKA. By trading off bandwidth for beams, we can have as a standard up to 8 widely 
separated field-of-views (FOVs) and optionally up to 244. 

{\bf Tied-array beams.} Within each of the station beams we can also
form multiple tied-array (TA) beams, applying proper phase delays between 
stations while adding them coherently. At the moment we can form TA
beams only using 12 Superterp sub-stations, as they 
share the single clock. Figure~\ref{ta127} shows the hexagonal pattern of 127 TA beams formed around the pulsar B2217+47. 
The signal-to-noise ratio of the other TA beams is about one order of magnitude smaller than the beam at the location
of the pulsar.
With 127 tied-array beams we can cover the whole station beam which is shown by the white circle. 
The FOV of the station beam is large enough to cover the entire Andromeda galaxy and then we can map it with TA beams 
in one single observation. Moreover, we can form a few station beams to cover a larger FOV and customize individual 
narrow TA beams pointing at different targets. The highest number of
TA beams formed in the commissioning 
observations so far was 217 (8 rings plus a central beam).

The list and description of all beam-formed modes that are well-tested and currently available to the wider community
can be found on the LOFAR web-pages\footnote{\tt http://astron.nl/radio-observatory/observing-capabilities/ depth-technical-information/major-observing-modes/beam-form}.
There are many possible modes or configurations, and the system is very flexible, to match different science goals. 
\begin{figure}[tbhp]
\begin{minipage}{\textwidth}
\begin{center}
\includegraphics[width=\textwidth]{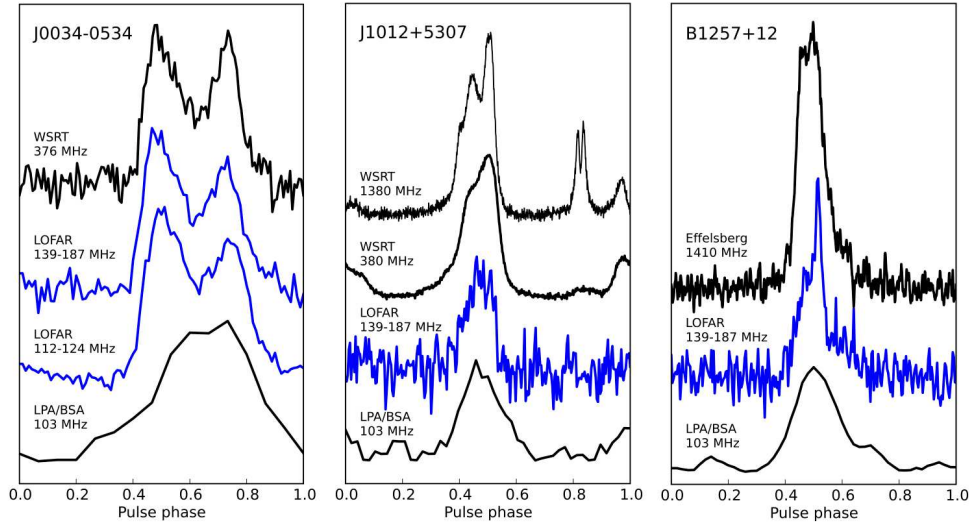}
\vskip -3mm
\caption[LOFAR MSP profiles]{LOFAR profiles observed in the HBA bands (blue) for the three millisecond pulsars J0034$-$0534, J1012+5307, 
and B1257+12 in comparison with profiles at 103~MHz with the BSA phased array in Puschino, and WSRT and
Effelsberg profiles at higher frequencies (black). Each bin in the LOFAR profiles is about $20~\mu$s wide. BSA profiles
are from the EPN pulsar database ({\tt http://www.jb.man.ac.uk/research/pulsar/Resources/epn/browser.html}).}
\label{msps}
\end{center}
\end{minipage}
\end{figure}
Different 
data products can be recorded, namely total intensity, full Stokes parameters, or complex voltage data.
All data are written in HDF5\footnote{\tt http://www.hdfgroup.org/HDF5/} format and we are already working to read it directly with 
{\tt DSPSR}\footnote{\tt http://dspsr.sourceforge.net} and 
{\tt PRESTO}\footnote{\tt http://www.cv.nrao.edu/\symbol{126}sransom/presto/} pulsar software.

{\bf RFI.} The RFI environment is very clean, much better than anticipated. The reasons for this
is that a) we are using 12-bit ADCs at the station level, so the
dynamic range is high; and b) the dipoles are located
very low to the ground and do  not pick up a much terrestrial interference. Typically, we flag 
about 1--2\% of data in HBA, and 3--4\% in LBA range. Below 30~MHz,
however, the data get very contaminated by RFI.

{\bf Full-core single-clock.} We are currently working on expanding
the number of stations that use the single clock, from six stations on the Superterp 
to the whole core of 24 stations within $\sim 1$~km radius. The work is ongoing and will be finished by the end
of October 2012. {\it This will further increase the raw sensitivity of the
  system by a factor 4!}

\section{LOFAR Highlights}
Here I present some of our recent pulsar results; some are  published or will be submitted
 soon. 

{\bf Pulsar timing.} LOFAR is very capable of, and we have already started doing, observations of 
millisecond pulsars (MSPs), as shown in Figure~\ref{msps}. LOFAR MSP profiles show a very high quality
at such a low frequency in comparison with previously acquired data using the Puschino BSA phased-array at 103~MHz.
These are the highest-quality detections of these pulsars ever made below 200~MHz.
\begin{figure}[tbhp]
\begin{center}
\includegraphics[width=0.7\textwidth]{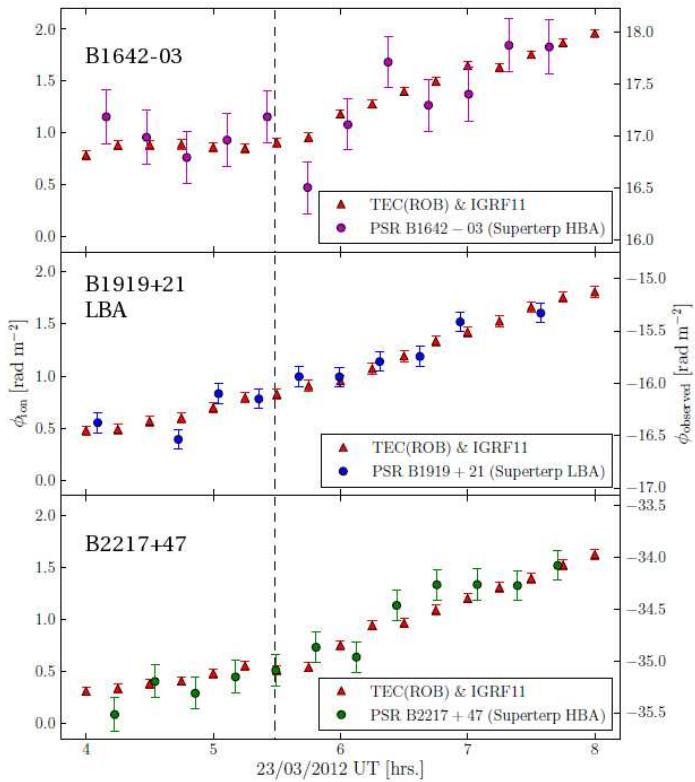}
\caption{Observed Faraday depths, $\phi_\mathrm{observed}$ (right axis), along the line-of-sights
toward the pulsars B1642$-$03 (top), B1919+21 (middle), and B2217+47 (bottom) vs. observing time.
Vertical dashed line designates the sunrise. The model output for each line-of-sight (red triangles)
is shown on the left axis (Figure taken from \cite[Sotomayor-Beltran et al. 2012]{ss+12}).}
\label{iono}
\end{center}
\end{figure}
For the MSP J0034$-$0534 comparison of LOFAR profiles with the 376-MHz WSRT profile shows that a small 
scattering tail becomes visible at lower frequencies (more apparent at 112--124~MHz), along with a slight 
change in the relative amplitudes of the two profile components.
We have already started timing observations of MSPs to test the system and pipeline, 
and with the full-core single-clock we will start the real campaign of timing MSPs. LOFAR pulsar timing
observations will be very important to get a handle on dispersion measure and pulse profile evolution
crucial for high-precision timing at high radio frequencies.

{\bf Ionospheric Faraday rotation calibration.} 
We started Faraday rotation monitoring to be able to measure
accurately pulsar rotation measures (RMs).  Figure~\ref{iono} (\cite[Sotomayor-Beltran et al. 2012]{ss+12}) presents
the observed Faraday depths for three pulsars together with the model
predictions based on the total electron content (TEC) maps from the
Royal Observatory of Belgium (ROB) and the International Geomagnetic
Reference Field (IGRF11). It can be clearly seen that our measurements
(circles) match the model (red triangles) very well. We are now
getting down to very robust and precise RM measurements of about
0.1~rad m$^{-2}$. The observations presented used only 1/6 of the
LOFAR's available bandwidth, thus showing a great potential for even
better RM measurements using the full bandwidth especially in the LBA
band.

{\bf Dispersion measure vs. Profile variations.}
\cite[Hassall et al. (2012)]{hsh+12} studied dispersion measure (DM) and profile variations
for four pulsars, B0329+54, B0809+74, B1133+16, and B1919+21 using simultaneous wideband observations
\begin{figure}[tbhp]
\begin{center}
\raisebox{8.3cm}{\includegraphics[width=0.6\textwidth,angle=270]{figs/0809cs.scr.ar-half.eps}}\includegraphics[width=0.35\textwidth]{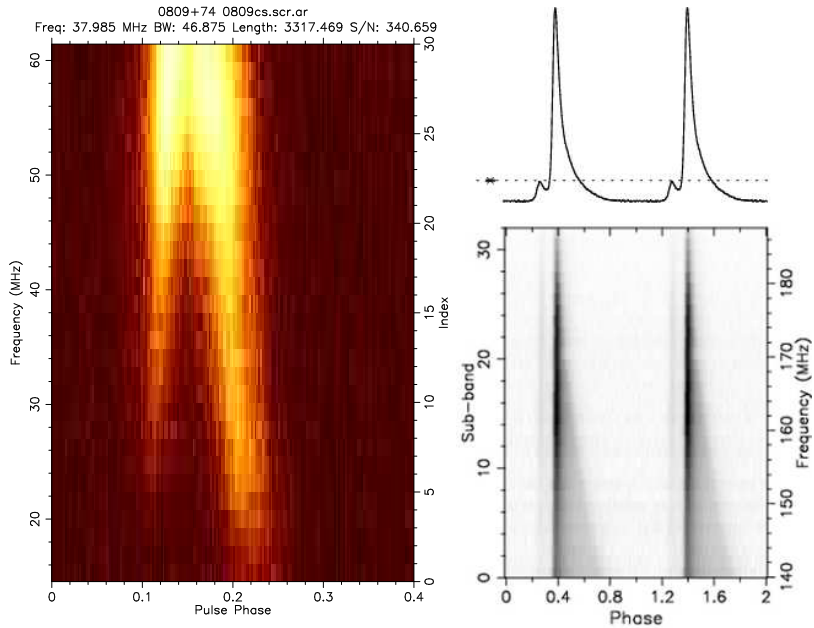}
\vskip 2mm
\caption{{\it Left:} LOFAR pulse average spectrum of PSR~B0809+74 at frequencies 15--62~MHz. {\it Right:}
Frequency-phase plot from one of the LOFAR HBA observations of PSR~B2111+46 with the average pulse profile
on the top. Grey scale designates the signal-to-noise ratio and remarkable evolution of the scattering tail
is clearly visible.}
\vskip -4mm
\label{0809-2111}
\end{center}
\end{figure}
with the LOFAR LBA and HBA at 40--190~MHz, the Lovell telescope at 1.4~GHz, and Effelsberg radio
telescope at 8~GHz. We found that the dispersion law is correct to better than 1 part in $10^5$ across
our observing band. We also put unique constraints on emission heights
for these pulsars using
aberration/retardation arguments and show that, for instance, in the case of the pulsar B1133+16 all radio
emission comes from a small region less than 59~km in altitude at a height of less than 110~km 
above the neutron star surface (only 0.2\% of the light cylinder). We found no evidence for the super-dispersion
delay previously reported at low frequencies (\cite[Shitov \& Malofeev 1985]{sm85}; \cite[Kuzmin 1986]{k86})
and suggest it could be caused by pulse profile evolution or a wrong fiducial point. We show that profile evolution
has a siginificant impact on high-precision pulsar timing and should be taken into account.

{\bf Low-frequency single-pulse studies.}
Figure~\ref{0809-2111} (left) shows the remarkable profile
evolution of the pulsar B0809+74 from 62 down to 15~MHz. We performed a thorough single-pulse analysis for
the pulsars B0809+74 and B1133+16 that show quite interesting results. For more details about single-pulse
studies of the pulsar B0809+74 see by Kondratiev et al. (these proceedings).

{\bf Pilot pulsar surveys.} We have already finished two pilot pulsar surveys with LOFAR. For more
details about the survey setup, search pipelines and results, see
Coenen et al. (these proceedings).

{\bf Low-frequency pulsar profiles.} Some of the examples of LOFAR pulsar profiles at HBA and LBA bands were
already shown in \cite[Stappers et al. (2011b)]{sha+11b}. Currently we have already detected more than 110
pulsars in the HBA and 12 pulsars in the LBA. We expect these numbers to significantly increase in the very
near future with the full-core single-clock, when the LOFAR raw sensitivity will be quadrupled. 
We are working on the ultimate LOFAR pulsar profile paper, and in particular on profile alignment with the
high-frequency WSRT and Jodrell Bank data.

{\bf Scattering studies.} 
LOFAR's low-frequency range and huge fractional bandwidth is ideal for pulsar scattering studies. 
Figure~\ref{0809-2111} (right) shows the benefits of the LOFAR's huge fractional bandwidth where you can see a remarkable 
scattering tail from the pulsar B2111+46 changing across the band. This allows us to study precisely the 
frequency dependency of scattering parameters of this and other pulsars.

\section{Future advancements}
LOFAR commissioning work is continuing and there are significant improvements which are coming by the end
of Fall 2012, namely:

\begin{itemize}
\item Expanding single clock to the full LOFAR core (end of October 2012). {\it This will quadruple LOFAR's
raw sensitivity.}
\item Reading HDF5\footnote{Hierarchical Data Format 5, {\tt http://www.hdfgroup.org/HDF5/}} data directly 
using {\tt DSPSR} and {\tt PRESTO} (nearly completed).
\item Doubling (almost) of the available bandwidth to about 80~MHz by implementing the 8-bit mode 
and potentially even 4-bit (end of October 2012).
\item Implementing online RFI excision (removal from raw data).
\item Creating sub-arrays and true Fly's Eye observations.
\end{itemize}

\section{Conclusions}

The results presented here have already proven the exceptional capabilities of the LOFAR
and opened up the whole new window of comprehensive studies of pulsars at low frequencies.
We have published first, intriguing results, with additional papers in
preparation. The forthcoming implementation of the full-core single-clock,
with the four-fold increase in sensitivity, will further enhance the
LOFAR pulsar capabilities.

~\\
{\bf Acknowledgements}\\
The LOFAR facilities in the Netherlands and other countries,
under different ownership, are operated through the International LOFAR
Telescope foundation (ILT) as an international observatory open to the global astronomical
community under a joint scientific policy. In the Netherlands, LOFAR
is funded through the BSIK program for interdisciplinary research and improvement
of the knowledge infrastructure. Additional funding is provided through
the European Regional Development Fund (EFRO) and the innovation program
EZ/KOMPAS of the Collaboration of the Northern Provinces (SNN). ASTRON
is part of the Netherlands Organization for Scientific Research (NWO).

\end{document}